\documentstyle[12pt]{article}

\begin{document}

\baselineskip=7mm
\renewcommand{\arraystretch}{1.3}

\newcommand{\cf}{{ f}}
\newcommand{\TeV}{\,{\rm TeV}}
\newcommand{\GeV}{\,{\rm GeV}}
\newcommand{\MeV}{\,{\rm MeV}}
\newcommand{\keV}{\,{\rm keV}}
\newcommand{\eV}{\,{\rm eV}}
\newcommand{\Tr}{{\rm Tr}\!}
\newcommand{\be}{\begin{equation}}
\newcommand{\ee}{\end{equation}}
\newcommand{\bea}{\begin{eqnarray}}
\newcommand{\eea}{\end{eqnarray}}
\newcommand{\ba}{\begin{array}}
\newcommand{\ea}{\end{array}}
\newcommand{\bmat}{\left(\ba}
\newcommand{\emat}{\ea\right)}
\newcommand{\refs}[1]{(\ref{#1})}
\newcommand{\ler}{\stackrel{\scriptstyle <}{\scriptstyle\sim}}
\newcommand{\ger}{\stackrel{\scriptstyle >}{\scriptstyle\sim}}
\newcommand{\lag}{\langle}
\newcommand{\rag}{\rangle}
\newcommand{\ns}{\normalsize}

\begin{titlepage}
\title{{\Large\bf Gauge Unification and Dynamical Supersymmetry Breaking}\\
                          \vspace{-4.5cm}
                          \hfill{\ns CBNU-TH 970530\\}
                          \hfill{\ns hep-ph/9705457\\}
                          \vspace{3.5cm} }
\author{ Eung Jin Chun,\hspace{.2cm}  
        Haewon Lee,\hspace{.2cm}  and \hspace{.2cm}  
        Won Sik l'Yi\\[.5cm]
  {\ns\it Department of Physics, Chungbuk National University}\\
  {\ns\it Cheongju, Chungbuk 360-763,  Korea} \\ 
        }
\date{}
\maketitle
\begin{abstract} \baselineskip=7.2mm {\ns
Under the assumption that all the gauge groups in supersymmetric 
theories unify at the fundamental scale, the numbers and the mass 
scales of messenger quarks and leptons, as well as the beta-function 
coefficient of the sector for dynamical supersymmetry breaking  
are constrained depending on various gauge mediation mechanisms.
For this, we use one-loop renormalization group equations and draw 
constraints on the scales in each  gauge mediation model.
\\[.3cm]
}\end{abstract}
\thispagestyle{empty}
\end{titlepage}

Dynamical supersymmetry breaking (DSB) \cite{witten} has been
known to occur in some supersymmetric gauge theories \cite{ads}.
Recently,  following the new understanding of the 
quantum behavior of supersymmetric gauge theories \cite{seiberg}, the number 
of theories for DSB increased rapidly \cite{skiba}.
Concurrently, various new mechanisms for transmitting supersymmetry breaking 
through gauge mediation have been proposed \cite{dnns}--\cite{mura}. 
The supersymmetric standard model (SSM) has to be implemented with
the sector for supersymmetry breaking and a way of its mediation 
to the SSM sector.  
It will be a future task to find a right DSB sector and messenger mechanism 
which can yield phenomenologically acceptable soft supersymmetry breaking 
in the SSM sector.

As is well-known, the minimal supersymmetric standard model supports the 
idea of grand unification \cite{unify}.  It remains however to be explained
why the grand unification scale $\sim 10^{16}$ GeV 
differs from the fundamental scale which could be the string scale
$\sim 10^{17-18}$ GeV \cite{kapl} or the Planck scale 
$\sim 10^{19}$ GeV.
This question has been addressed in SSM-like string theories \cite{string} 
where the masses of extra fermiones can reside anywhere between 
the electroweak scale and the string scale.
It can be expected that this problem is resolved in gauge mediation models of 
supersymmetry breaking where  extra heavy quarks and leptons are 
necessary and their mass scales  are determined from the successful
prediction of ordinary superparticle masses.
The existence of such messenger fermions may remove the discordance between 
the conventional unification scale and the fundamental scale. 
Then the fundamental theory should be  such that 
not only the standard model gauge interactions  but also DSB gauge
interactions unify at the fundamental scale of the theory.
In this case, the supersymmetry breaking scale can be also determined 
dynamically in terms of the fundamental scale.
An attempt to find a realistic string model with such a property was made in 
Ref.~\cite{fara}.
The aim of this paper is to investigate the general 
consequences of the ultimate unification of the SSM sector and the DSB sector.
Specifically, we will draw restrictions on the various scales in the theory
and the number of messenger quarks and leptons depending on theories for 
DSB and mechanisms for gauge mediation.

\bigskip

We begin with considering the renormalization group equations
of the SSM and DSB gauge coupling constants.   Our analysis will rely on
a rough order-of-magnitude calculation which is enough for our purpose.
Gauge mediation models contain additional vector-like quarks or 
leptons (denoted by $f, \bar{f}$) whose masses (denoted by $M_m$) 
can be generated  by the following schematic form of superpotential:
\be \label{sff}
 W = \lambda S f \bar{f} \,.
\ee
Here the field $S$ can be a fundamental or a  higher dimensional 
composite field.
The nonzero vacuum expectation values (VEVs) of $S$ and its F-term $F_S$ 
result from the mediation of supersymmetry breaking in the DSB sector.
Then the ordinary superparticles obtain  the soft masses 
of the order ${\alpha\over 4\pi}\Lambda_S$ where 
\be  \label{ls}
  \Lambda_S \approx {F_S \over S} \approx  (10 \sim 100) \TeV
\ee
and $\alpha$ is a standard model fine structure constant.
The messenger quarks and leptons at the messenger scale $M_m$
participate in the renormalization group evolution up to 
the fundamental scale $M_X$ at which all gauge groups unify.
The assumption of the gauge unification allows us to 
compute the dynamical scale $\Lambda_D$ once the gauge structure of the 
DSB sector (more precisely the coefficient of the $\beta$-function) is 
fixed.  For simplicity, we assume that there is only one dynamical scale in
the DSB sector.  This assumption is not so restrictive since the largest
dynamical scale can be taken when the DSB sector has a product group.

The one-loop renormalization group evolution of the gauge couplings is
given by
\bea \label{rge}
 \alpha_X^{-1} &=& \alpha_i^{-1}(M_Z)+ {b_i\over 2\pi}\ln{M_X\over M_Z}
                    - {n_i\over2\pi}\ln{M_X\over M_m} \nonumber\\
              % &=& \alpha_3^{-1}(M_Z)+ {b_3\over 2\pi}\ln{M_X\over M_Z}
              %      - {n_3\over2\pi}\ln{M_X\over M_m} \nonumber\\
               &=& {b\over 2\pi}\ln{M_X\over \Lambda_D} 
\eea
where $b_{i}$ is  the minimal value of the coefficient of the one-loop
$\beta$-function ($b_1= -33/5, b_2=-1, b_3=3$), 
and $-n_{i}$ is the contribution from the messenger fermions 
at the mass scale $M_m$, and $b$ is the 
coefficient of the DSB sector. 
%For instance, 
%when the messenger quarks $N_{32}\times
%[({\bf 3},{\bf 2})+(\bar{\bf 3},\bar{\bf 2})]$, $N_3 \times [({\bf 3},{\bf 1})
%+(\bar{\bf 3},{\bf 1})]$ and leptons 
%$N_2 \times [({\bf 1},{\bf 2})+({\bf 1},\bar{\bf 2})]$ are added,  
%we have $n_2=3N_{32}+N_2$ and $n_3=2N_{32}+N_3$.
Note that the extra quarks or leptons do not have to form  complete
multiplets of a unification group, e.g, $({\bf 5}+ {\bf\bar{5}})$ or 
$({\bf 10}+{\bf\bar{10}})$ of $SU(5)$ as also discussed  in other studies
\cite{fara,martin}.  
%The numbers of SSM-like representations
%were constrained in Ref.~\cite{martin}
%under the condition of perturbative gauge couplings but
%without assuming unification.
%In Eq.~\refs{rge}, we omitted the evolution equation for the $U(1)_Y$ gauge
%coupling constant which should also contain the threshold effect at $M_m$.
%In our study, messenger fermions are assumed to carry appropriate $U(1)_Y$
%charges consistent with unification.
With the simple one-loop renormalization group equation \refs{rge}, we can 
draw information on the values of $n_i$ and $b$ which are compatible 
with the unification idea in various gauge mediation models.
In our discussion, we ignore the two-loop evolution which involves also
information on the messenger fermion Yukawa couplings and two-loop
$\beta$-function of the DSB sector.   We  expect that two-loop effects
cause no essential change in the prediction of the numbers $n_i$ 
and $b$ and the orders of magnitude of various scales.
More precise  phenomenological discussions  at  two-loop order 
in gauge mediation models with conventional unification group $SU(5)$ 
have been performed in Refs.~\cite{bagger}. 

First, we get the relation between the messenger scale and the
unification scale $M_X$,
\be \label{Mm}
   M_m = M_X \left( M_U \over M_X \right)^{4/n}\,, \qquad
   M_U \equiv M_Z e^{2\pi(\alpha_2^{-1}-\alpha_3^{-1})/4}
\ee
where $n\equiv n_3-n_2$.
Here $M_U$ is the usual unification scale $\approx 2.4\times 10^{16}$ GeV.
For the calculation, we use the central values; $\alpha_1=1/58.97,
\alpha_2=1/29.61, \alpha_3=0.118$ at the scale $M_Z$ \cite{data}.  
Assuming $M_X>M_U$ to ensure the experimental bounds on the proton
lifetime are not violated, $n$ must be positive or zero.  
When $n=0$, the messenger scale $M_m$
is not related to the unification scale, and $M_X = M_U$.
The messenger scale and the DSB scale are related by the equation,
\be \label{n0}
  2\pi\alpha_U^{-1} = n_2 \ln{M_U\over M_m} + b \ln{M_U\over \Lambda_D} \,,
\ee
where $\alpha_U \approx 1/24$ is the usual unification coupling constant.
%This is the case when the messenger quarks and leptons form  complete
%multiplets of the unified group. 
On the contrary to the case with $n=0$,
the unification scale can be pushed up when $n\geq 1$.
The number $n$ cannot be arbitrarily large. The upper bound $M_m<10^{16}$ GeV
[see below Eq.~\refs{hii}] implies $n\leq 3$ as can be seen from Eq.~\refs{Mm}.
There is also a lower bound on the messenger scale which 
becomes smaller for a larger $M_X$ and a smaller $n$.  The smallest 
messenger mass  can be obtained by taking $n=1$ and $M_X=M_{pl} \approx
1.2\times10^{19}$ GeV: that is, $M_m>2\times 10^8$ GeV.  
Of course, this lower bound on $M_m$ is not applied to the case with $n=0$
[see Eq.~\refs{n0}].
Note that the messenger scale would be related to the axion scale or 
the heavy right-handed neutrino scale when $M_m \approx 10^{10} 
\sim 10^{12}$ GeV, which can be obtained  with $n=1$ and $M_X \approx
3\times10^{18} \sim 7\times10^{17}$ GeV.
Given $n$ (or $n_3$) and $n_2$, the number $n_1$ is constrained 
by the well-known relation
$$ {n_1-n_2 \over n_3-n_2 } = {b_1-b_2 \over b_3-b_2} $$
from which one gets,
\be \label{n1}
 n_1 = n_2 - {7\over 5} n  \,.
\ee
If $n=0$, it is required that $M_X = M_U$ and $n_1=n_2=n_3$, 
which is the case when the messenger fermions form  complete representations 
of certain unification group.
The number $n_1$ depends upon the $U(1)_Y$ charge assignment to the messenger 
fermions.  Later, we will see how the relation \refs{n1} restricts the
number of SSM-like particles with the standard $U(1)_Y$ charges.
We are now ready to discuss implications of the ultimate unification
in various types of gauge mediation models which
can be classified essentially into two classes: models with 
indirect, or direct mediation. 

\bigskip

(I) {\bf Indirect mediation models} :  In this class of models, 
supersymmetry breaking in the DSB sector is transmitted first to 
the messenger quarks and leptons by an intermediate 
gauge interaction and then to the SSM sector as described above.  
As a consequence, the messenger fermions get masses of order 
$M_m \approx \lag S \rag \approx \alpha' \Lambda_D/4\pi$ where $\alpha'$ is 
the intermediate gauge coupling constant.
If one consider renormalizable interactions only, the supersymmetry 
breaking scale of the DSB sector is given by $\sqrt{F} \approx \Lambda_D$.  
In general, the gravitino mass is $m_{3/2} \approx F/M_P$ where
$M_P\approx 2.4\times 10^{18}$ GeV is the reduced
Planck mass.   Generic supergravity contribution to the soft masses of the
superparticles (proportional to $m_{3/2}$) are not favor-blind and thus 
generate too large FCNC effects.
To avoid this, we require a conservative constraint that the gravitino mass 
is smaller than the typical soft mass: $m_{3/2} < 100$ GeV, 
that is, $\sqrt{F} <10^{10}$ GeV.  
From the fact that $\Lambda_S\approx 10$ to 100 TeV and that $M_m$ is roughly
of the same order or larger than $\Lambda_S$ \cite{dnns}, one can draw the
limit: $10^4 \GeV < M_m < 10^7 \GeV$.  
Now that the hierarchy between $M_m$ and $\Lambda_D$ 
is generated by loop effects, the more realistic 
bound on $\Lambda_D$ is  $10^5 \GeV < \Lambda_D < 10^8 \GeV$.
Altogether, we get the hierarchy among the  scales:
\be \label{hi}
   \Lambda_S \ler  M_m < \Lambda_D < 10^{8} \GeV \,.
\ee
Note here that the gravitino mass ($m_{3/2} \approx \Lambda_D^2/M_P$) is 
in the range $4 \eV< m_{3/2} < 4 \MeV$.  Such a light gravitino can 
yield distinctive phenomenological consequences in cosmology \cite{ckk}
and collider physics \cite{kkk}.

The upper bound on $\Lambda_D$ can be relaxed  if the DSB sector contains
nonrenormalizable terms.
The interplay between two  scales $\Lambda_D$ and $M_P$ in the effective
DSB superpotential can give rise to a VEV $v$ larger than $\Lambda_D$: 
$v \approx M_P(\Lambda_D/M_P)^\kappa$ with 
$0<\kappa<1$ \cite{dnns}. 
Now, $F/v$ in the DSB sector plays a role of $\Lambda_D$ in
Eq.~\refs{hi}. The order of the  DSB superpotential is 
\be \label{Wnon}
  W \approx {v^{m+3} \over M_P^m}
\ee 
when nonrenormalizable terms of dimension $m+3$ give the largest
contribution. Then  the supersymmetry breaking scale is given by
$F \approx v^{m+2}/M_P^m$, and hence $F/v \approx v^{m+1}/M_P^m$.
Assuming the supersymmetry breaking scale $\sqrt{F}$ is below 
the DSB scale, we obtain
\be \label{hii}
 \Lambda_S \ler  M_m < {F\over v} < \sqrt{F} < \Lambda_D < v <10^{16} \GeV\,.
\ee
Here the upper bound $v<10^{16} \GeV$ comes  from 
$v \Lambda_S  < F<10^{20} \GeV^2$. 
In this case, the natural range of the DSB scale can be inferred to be
$10^6 \GeV < \Lambda_D < 10^{16} \GeV$.  

\medskip

As shown, the messenger mass in indirect mediation models has an upper
bound $M_m < 10^7$ GeV which is below the smallest value $M_m \approx
2\times 10^8$ GeV in case of $n\neq 0$.  Therefore, indirect mediation models
can only employ $n=0$, and thus $M_X=M_U$.  In this case, the condition
$n_1=n_2=n_3$ has to be satisfied. A trivial way to achieve such a condition
is to take the charge assignments such that the messengers form  complete
representations of a unification group as noted before.
For renormalizable indirect mediation models, the allowed ranges of the
messenger scale and the DSB scale [discussed above Eq.~\refs{hi}] can be 
realized for a quite restricted ranges of $n_2$ and $b$.  From Eq.~\refs{n0},
one finds,
\be  \label{first}
  1\leq  n_2 \leq 5 \,,\qquad   1 \leq b \leq 7 \,.
\ee
Here the lower limits are trivial and 
the upper limit on $n_2$ comes from the perturbative unification 
condition $\alpha_X < 1/3$.
In fact, two numbers $n_2, b$ are correlated and roughly speaking,
$n_2+b=6,7,8$ has to be satisfied.
This result is consistent with Dubovsky et.al.\ \cite{dubo}.  
A difference results from the fact that  
the unification of all gauge sectors at once is assumed and 
the change of $\alpha_X$ due to non-zero $n_2=n_3$ is taken into account 
in our case. 
Indeed, in the case of $n=0$,  identification of the unification scale $M_X$
with, e.g., $M_{pl}$ instead of $M_U$ is possible 
if there is two step unification; the SSM gauge sector unifies 
at $M_U$ and then the ultimate unification including the DSB sector occurs 
somewhere between $M_U$ and $M_{pl}$ as considered in Ref.~\cite{dubo}.
This scheme then requires an explanation 
how the scales different from, e.g., $M_{pl}$ can be generated.

To summarize, renormalizable indirect mediation models require a DSB sector 
with a small $b \leq 7$ to achieve unification.  
The number of DSB models with such a small $b$ in the literature
is very limited.
To our knowledge, there are only a few models with 
$b<10$.  They are $SU(4)\times SU(3) \times U(1)$ model with $b=8$
\cite{crs}, and the models listed in Ref.~\cite{dubo}.
In order to construct  more DSB models  with a small $b$, 
one would need to find a way to use higher dimensional 
representations of a given DSB gauge group.
For nonrenormalizable class models where $\Lambda_D < 10^{16}$
GeV, essentially no upper limit on $b$ ($b <175$) can be drawn.

\bigskip

(II) {\bf Direct mediation models} :  
The above conclusion can change a lot in this type of models 
where the value of $M_m$ (or $\Lambda_D$) can be larger.
In this scheme, the field $S$ belongs 
directly to the DSB sector and thus $F_S=F$.
As in the case of indirect mediation, there could be renormalizable and
nonrenormalizable classes of models.  
In nonrenormalizable class of models,
a higher dimensional DSB superpotential as in Eq.~\refs{Wnon} generates
a large VEV of $S$.
On the other hand, since the dimension of the field $S$ can be
one or bigger, the messenger mass is in general given by 
\be 
M_m \approx \lag S \rag \approx M_P(v/M_P)^d
\ee
where $d$ is the dimension of the operator $S$.
The case with $d=1$ is explored in Ref.~\cite{amm}, and a model with a 
composite field $S$ ($d>1$) is presented in Ref.~\cite{hmm}.
In the former case, there could be other messenger 
quarks or leptons whose masses can be much smaller than $\Lambda_S$. 
These light messenger fields are
known to drive the soft mass squared of the ordinary squarks to a negative
value \cite{hmm,pt2}.  Therefore, we assume the models without such a light
messenger fermion.  In these cases, the mass scale $\Lambda_S$ is given by 
$\Lambda_S \approx F/v$.  Therefore, we have
\be \label{hiii}
  \Lambda_S \approx {F\over v} < \sqrt{F} < \Lambda_D < v <10^{16} \GeV \,.
\ee
Recalling $F/v \approx M_P(v/M_P)^{m+1}$ [see below Eq.~\refs{Wnon}],
the bound $v<10^{16}$ GeV puts a limit: $m\leq 5$.
Furthermore, the fact that the messenger mass 
($M_m \approx M_P(\Lambda_S/M_P)^{d/m+1}$)
must be heavier than around 100 GeV restricts $d$: $d\leq m+1$.

A large VEV $v$ can be obtained dynamically also  
in the renormalizable class of models.
In this case, one consider an one-loop effective scalar potential of 
the form $V=f(v)\Lambda_D^4$ where $f(v)$ is a function of a field with 
a VEV $v$.  Then, the function $f(v)$ may be minimized at 
a large $v>\Lambda_D$ \cite{mura}.
In these models, we obtain 
\be \label{hiv}
 \Lambda_S \approx {\Lambda_D^2 \over v} < \sqrt{F}\approx \Lambda_D < v 
   < 10^{16} \GeV \,.
\ee
Here the messenger scale is given by $M_m \approx v$ and 
$\Lambda_D<10^{10}$ GeV as above.

\medskip

Let us now extract rough constraints on the numbers $n, n_2$ and $b$ 
for each class of indirect mediation models.
To be specific, let us take two canonical candidates of $M_X$: 
the Planck scale $M_{pl}$
and the string scale $M_{st} \approx 5\times10^{17}$ GeV \cite{kapl}. 
For $n=0$, the discussion in part (I) applies here as well.
However, we will concentrate on the cases with $n \geq 1$ 
for which  the unification scale $M_X$ can be made
close to the Planck scale.
As we discussed, given $M_X$ and $n$ determines the messenger scale 
residing in the range: $M_m \approx 2\times 10^{8} \sim 10^{16}$ GeV.
Furthermore, putting $M_m = \Lambda_D<10^{16}$ GeV in Eq.~\refs{rge}
one finds the upper bound: 
$n_2+b<(2\pi\alpha_U^{-1}-\ln(M_X/M_U))/\ln(M_X/10^{16} \GeV) \approx
20\,  (38)$ for $M_X=M_{pl}\, (M_{st})$.
The number $n_2$ is constrained individually 
assuming a perturbative unification:
$n_2/n < 2\pi(\alpha_U^{-1}-\alpha^{-1}_{per})/4\ln(M_X/M_U) - 1/4$.
Taking $\alpha_{per}=1/3$, we get  $n_2/n<5.8\, (10.8)$ 
for $M_X=M_{pl}\, (M_{st})$.

\begin{table}
\caption{ Various scales in unit of GeV
and allowed numbers of $n,n_2$ and $b$ in the 
nonrenormalizable class of models.
The messenger mass $M_m$ for each $n$ is the same as in Table 2.} 
\begin{tabular}{|c|c|c|c|c|c|c|c|}
\multicolumn{8}{l}  {For $M_X=M_{pl}$.} \hfill \\ \hline 
$n$ &  $(d,m)$ & $v$ & $\sqrt{F}$ & $n_2$ & $b$ & $\Lambda_D$ & 
$\Lambda_S$\\ \hline
1 &  (3,3) & $1\times10^{15}$ & $9\times10^9$ &
  $5 \sim  0$ & $2\sim15$  & 
  $1\times10^{10}\sim1\times10^{15}$  & $8\times10^4$\\
2 & (1,2) & $5\times10^{13}$ & $1\times10^9$ &
  $10\sim0$ & $1\sim11$  & 
  $2\times10^9 \sim 2\times10^{13}$ &  $2\times10^4$\\
3 & (1,4) & $3\times10^{15}$ & $5\times10^9$ &
  $12\sim0$ & $3 \sim 17$  & 
  $9\times10^9 \sim 2\times10^{15}$ & $7\times10^3$ \\ \hline 
\multicolumn{8}{l}  {For $M_X=M_{st}$.} \hfill \\ \hline 
2 & (1,3) & $1\times10^{15}$ & $1\times10^{10}$ &
  $ 12\sim0$ & $5 \sim 24$  & 
  $2\times10^{10} \sim 1\times10^{15}$ & $1\times10^5$ \\
3 & (1,5) & $9\times10^{15}$ & $7\times10^9$ &
  $12\sim0$  & $6 \sim 36$  & 
  $8\times10^9 \sim 8\times10^{15}$ & $5\times10^3$  \\ \hline
\end{tabular}
\end{table}

Given specific models, only certain combinations of the numbers 
$n_2$ and $b$ can be consistent with unification
which can be seen from Eq.~\refs{rge} together with Eqs.~\refs{hiii} and
\refs{hiv}.  For the nonrenormalizable class of models,  
the dimensionalities $(d,m)$ are also constrained due to
the relations: $M_m \approx M_U(M_X/M_U)^{n-4 \over n}$, 
$v \approx M_P(M_m/M_P)^{1/d}$ and $\Lambda_S \approx M_P(v/M_P)^{m+1}$
implying that fixing $n,d$ and $m$ determines the values $v$ and $\Lambda_S$
for each $M_X$. 
First, the number $d$ is restricted by the bound $v<10^{16}$ GeV:
\be \label{dis}
 d< { \ln{M_U\over M_P} + {4-n\over n} \ln{M_U \over M_X} \over
     \ln{10^{16}\GeV \over M_P} } \,.
\ee
This shows that the bound on $d$ becomes larger  for 
a larger $M_X$ and a  smaller $n$.   When  $n=1$, one has $d\leq 4 \,(2)$
for $M_X=M_{pl}\,(M_{st})$. On the other hand, only $d=1$
can be compatible for $n=2,3$.
The integer $m$ is constrained by the bound on $\Lambda_S \approx
M_P(M_m/M_P)^{m+1 \over d} \approx 10^4\sim10^5$ GeV:
\be  \label{mis}
 {m+1 \over d} \approx { \ln{\Lambda_S\over M_P} \over 
   \ln\left({M_U\over M_P}\right) + 
  {4-n\over n} \ln\left(M_U\over M_X\right) }\,.
\ee
The integer pairs $(d,m)$ most closely satisfying this relation are shown
in the second column of Table 1.
Now, the restricted ranges of $n_2$ and $b$ 
can be obtained from the constraints:
$\sqrt{F} < \Lambda_D < v$ and $\sqrt{F} \approx M_P (v/M_P)^{1+m/2} 
<10^{10}$ GeV where $\Lambda_D$ is obtained by equating the first and 
the third line of Eq.~\refs{rge}.  In selecting $n_2,b$, 
we impose also the perturbative unification: $\alpha_X < 1/3$.
The allowed combinations of $(d,m)$ and $(n_2,b)$ for given $n$ and 
$M_X$ together with the corresponding supersymmetry breaking
scale $\sqrt{F}$ and dynamical scale $\Lambda_D$ are presented  in Table 1.  
As one can see, the allowed values of $b$ can be 
very large and becomes larger for a  smaller $M_X$.  The number $n_2$ is
restricted to a small value when $b$ is large, and vice versa, 
as in part (I) [see below Eq.~\refs{first}].
Therefore, most DSB models using the nonrenormalizable direct
mediation mechanism can be consistent with the idea of unification.  
Note, however, that the supersymmetry breaking scale
tends to be large ($\sqrt{F} >10^9$ GeV), and becomes larger for a 
smaller $M_X$.  Such a large scale gives rise to a heavy gravitino;
$m_{3/2} > 0.4$ GeV which is dangerous cosmologically \cite{ellis}.
However, such a heavy gravitino could be diluted away by a 
late-time entropy production, e.g., thermal inflation \cite{lyth}.

\begin{table}
\caption{ Various scales in unit of GeV 
         and allowed numbers of $n,n_2$ and $b$ in the 
          renormalizable class of models.} 
\begin{center}
\begin{tabular}{|c|c|c|c|c|c|}
\multicolumn{6}{l} {For $M_X=M_{pl}$.}  \\  \hline
$n$ & $M_m$ & $n_2$ & $b$ & $\Lambda_D$ & $\Lambda_S$ \\ \hline
1 & $2\times10^8$ &  0 & 5 & 
 $2\times10^6$  & $3\times10^4$ \\
2 & $5\times10^{13}$ & $10\sim1$ &  $1\sim 6$ & 
  $7\times10^8 \sim 4\times10^9$  & $1\times10^4 \sim 3\times10^5$ \\
3 & $3\times10^{15}$ & $10\sim0$ & $3\sim7$ & 
 $5\times10^9 \sim 1\times10^{10}$ & $7\times10^3 \sim 3\times10^{4}$ \\ 
\hline
\multicolumn{6}{l} {For $M_X=M_{st}$.}  \\ \hline 
1 & $3\times10^{12}$ & $9\sim0$ & $2\sim7$ & 
   $2\times10^8 \sim 9\times10^8$ & $2\times10^4 \sim 3\times10^5$ \\
2 & $1\times10^{15}$ & $ 10\sim0$ & $5\sim8$ & 
   $3\times10^9 \sim 9\times10^9$ & $7\times10^3 \sim 8\times10^4$ \\
3 & $9\times10^{15}$ &  6 & 7 & 
   $8\times10^{9}$ & $8\times10^3$ \\ \hline
\end{tabular}
\end{center}
\end{table}
In the renormalizable class of models, the upper bounds on $n_2$ and  $b$
are more restrictive because of $\Lambda_D <10^{10}$ GeV.
Following the above process now with the relation, 
$\Lambda_S \approx \Lambda_D^2/M_m$, 
we can get constraints on $n_2$ and $b$ which 
are summarized in Table 2. Contrary to the nonrenormalizable models,
a large  $b\geq 9$ is not permitted.  On the other hand, 
the supersymmetry breaking 
scale can be as small as $2\times10^6$ GeV implying a light gravitino
($m_{3/2} \approx 2$ keV) which can form warm dark matter.

%The field $S$ could be a nonrenormalizable operator 
%(i.e., $M_m \approx M_P(v/M_P)^d <v$), but our results do not change even in
%this case since the condition $\Lambda_D<v$ is not used.

Let us finally discuss how the relation \refs{n1} constrains more 
the content of messengers.  To see this explicitly, 
we take SSM-like particles for the messengers: that is, 
$N_{32}\times [({\bf 3},{\bf 2})+(\bar{\bf 3},\bar{\bf 2})]_{1/6}$, 
$N_3 \times [({\bf 3},{\bf 1}) +(\bar{\bf 3},{\bf 1})]_{2/3}$,
$N'_3 \times [({\bf 3},{\bf 1}) +(\bar{\bf 3},{\bf 1})]_{1/3}$,
$N_2 \times [({\bf 1},{\bf 2})+({\bf 1},\bar{\bf 2})]_{1/2}$,  
and $ N_1 \times [({\bf 1},{\bf 1})+({\bf 1},{\bf 1})]_{1}$ are added.
Here the subscripts denote the absolute values of the $U(1)_Y$ charges.
It is trivial to calculate the numbers $n_i$ in this case;
\bea \label{trivial}
 n_1 &=& {6\over 5} [ {1\over6} N_{32} +{4\over3} N_3 + {1\over3} N'_3
         + {1\over2} N_2 + N_1 ] \nonumber \\
 n_2 &=& 3 N_{32} + N_2 \\
 n_3 &=& 2 N_{32} + N_3 + N'_3  \,. \nonumber
\eea
Then the relation \refs{n1} tells us that any set of integers $N_3, N_1
, N_{32}$ and $n$  satisfying the equation; 
\be
  N_3+ N_1 -2 N_{32} +{3\over2} n =0 \,, 
\ee
is compatible with the unification.
Therefore, in  the case of SSM-like messengers, the unification of DSB sector 
with the SSM sector can be achieved if the messenger contents with 
$n=2$ and $N_3+N_1-2N_{32}=-3$ are taken.
It is straightforward to generalize this argument to the cases with
any exotic selection of messenger contents.

\bigskip

In conclusion, we investigated the consequences of the assumption that 
the weak scale and the supersymmetry breaking scale are generated 
dynamically from the scale of a fundamental theory, $M_X$,
 at which gauge couplings
of the supersymmetric standard model and the supersymmetry breaking sector 
unify.  Considering one-loop renormalization group evolution, the
number and the mass scale of extra vector-like quarks and leptons (messenger
fermions), and the structure of the dynamical supersymmetry breaking sector
are constrained  depending on the ways of mediating supersymmetry breaking.

In indirect mediation models where the messenger fermion masses are smaller 
than about $10^7$ GeV, the unification scale can not be changed from 
the usual grand unification value $\approx 2\times 10^{16}$ GeV.
In its renormalizable class of models 
the one-loop $\beta$-function coefficient of the DSB sector $b$ has to be
less than 8 corresponding to the condition of the DSB scale 
$\Lambda_D < 10^8$ GeV. 
We notified that there are only a few known examples with such a small $b$.
For the nonrenormalizable class of models, no bound on $b$ can be found.

In direct mediation models, the messenger mass can be larger than about 
$2\times10^8$ GeV but smaller than about $10^{16}$ GeV.
Furthermore, the mismatch between the usual unification scale 
and e.g., the Planck scale can be removed.  
For this, we need a kind of doublet-triplet splitting for messenger quarks 
and leptons with a small difference between the numbers of 
triplets and doublets ($n=n_3-n_2= 1,2,3$).  
A large value of the messenger mass is obtained when the fields in the
DSB sector get large vacuum expectation values ($<10^{16}$ GeV) 
by the presence of nonrenormalizable terms or by a loop-improved effective 
scalar potential.  

In the former case, phenomenologically acceptable models are shown to be
compatible with a large $b\leq 36$ or $n_2 \leq 12$ 
and with a DSB scale $\Lambda_D$ in the
range: $2\times 10^9 \sim 8\times 10^{15}$ GeV.  The supersymmetry breaking
scale $\sqrt{F}$ tends to be large indicating a large mass of
the gravitino: $m_{3/2} > 0.4$ GeV.  This heavy gravitino may cause 
cosmological troubles unless some dilution mechanism by 
a late-time entropy production takes place.
In the latter case, due to the restriction $\sqrt{F} \approx \Lambda_D <
10^{10}$ GeV, a small $b$ is acceptable: $b\leq8$,  similarly to the indirect
renormalizable models. But $n_2$ can be large: $n_2 \leq 10$.
The gravitino can be as light as 2 keV to form dark matter.
As noted, most supersymmetric gauge theories exhibiting  dynamical 
supersymmetry breaking in the literature
have $b$ larger than 10, and thus can be used only 
in a unified theory with the nonrenormalizable direct mediation mechanism.

\bigskip

{\bf Acknowledgment}: This work is supported by Non Directed Research
Fund of Korea Research Foundation, 1996.  E.J.C. is a Brain-Pool fellow.


\begin{thebibliography}{30}
\bibitem{witten} E. Witten Nucl.\ Phys.\ B 188, 513 (1981).
\bibitem{ads} I. Affleck, M. Dine and N. Seiberg, Phys.\ Lett.\ B137, 187
        (1984); Phys.\ Rev.\ Lett.\ 52, 1677 (1984); 
        Phys.\ Lett.\ B140, 59 (1984); Nucl.\ Phys.\ B256, 557 (1985); 
        D. Amati, K. Konishi, Y. Meurice, G. C. Rossi and G. Veneziano, 
        Phys.\ Rep.\ 162, 169 (1987).
\bibitem{seiberg} N. Seiberg, Phys.\ Rev.\ D49, 6857 (1994); 
         Nucl.\ Phys.\ B435, 129 (1995); for  reviews, see, 
         K. Intriligator and N. Seiberg, hep-th/9509066; 
         M. Peskin, hep-th/9702094.
\bibitem{skiba} For a review and references, see, W. Skiba, hep-th/9703159.
\bibitem{dnns}  M. Dine and A. Nelson, Phys.\ Rev.\ D48, 1277 (1993); 
         M. Dine, A. Nelson and Y. Shirman, Phys.\ Rev.\ D51, 1362 (1995);
         M. Dine, A. Nelson, Y. Nir and Y. Shirman, Phys.\ Rev.\ D53, 2658
         (1996);
\bibitem{amm} E. Poppitz and S. P. Trivedi, hep-ph/9609529;
         N. Arkani-Hamed, J. March-Russell and H. Murayama, hep-ph/9701286.
\bibitem{rmnr}  L. Randall, hep-ph/9612426; 
        R. N. Mohapatra and S. Nandi, hep-ph/9702291;
        S. Raby, hep-ph/9702299.
\bibitem{hmm} N. Haba, N. Maru and T Matsuoka, hep-ph/9703250;
         Y. Shadmi, hep-ph/9703312.
\bibitem{pt2} E. Poppitz and S. P. Trivedi, hep-ph/9703246;
\bibitem{mura} E. Witten, Phys.\ Lett.\ 105B, 267 (1981); 
          H. Murayama, preprint LBNL-40286, hep-th/9705271.
\bibitem{unify} U. Amaldi, W. de Boer and H. F\"urstenau, Phys.\ Lett.\ B260, 
         447 (1991); C. Giunti, C. W. Kim and U. W. Lee, Mod.\ Phys.\ Lett.\
         A6, 1745 (1991); J. Ellis, S. Kelley and D. V. Nanopoulos, Phys.\
         Lett.\ B260, 131 (1991); P. Langacker and M. X. Luo, Phys.\ Rev.\
         D44, 817 (1991).
\bibitem{kapl} V. S. Kaplunovsky, Nucl.\ Phys.\ B307, 145 (1988); Erratum,
              {\it ibid} B382, 436 (1992).
\bibitem{string} I. Antoniadis, J. Ellis, S. Kelley and D. V. Nanopoulos, 
        Phys.\ Lett.\ B272, 31 (1991); S. Kelley, J. Lopez, and D. V.
        Nanopoulos, Phys.\ Lett.\ B278, 140 (1992); M. K. Gaillard and R. Xiu, 
        Phys.\ Lett.\ B296, 71 (1992), S. P. Martin and P. Ramond, Phys.\
        Rev.\ D51, 6515 (1995); for a review, see K. R. Dienes, 
        to appear in Phys.\ Rept.\ (hep-ph/9702045).
\bibitem{fara} A. E. Farragi, Phys.\ Lett.\ B387, 975 (1996).
\bibitem{martin} S. P. Martin, Phys.\ Rev.\ D55, 3177 (1997).
\bibitem{bagger} C. D. Carone and H. Murayama,  Phys.\ Rev.\ D53, 1658 (1996);
         J. A. Bagger, K. T. Matchev, D. M. Pierce and R. Zhang,
         Phys.\ Rev.\ Lett.\ 78, 1002 (1997).
\bibitem{data} Particle Data Group Review of Particle Properties, Phys.\ Rev.\
           D54, 1 (1996).
\bibitem{ckk} E. J. Chun, H. B. Kim and J. E. Kim, Phys.\ Rev.\ Lett.\
        72, 1956 (1994);
        A. de Gouvea, T. Moroi and H. Murayama, Phys.\ Rev.\ D56,
        1281 (1997). 
\bibitem{kkk} S. Dimopoulos, M. Dine, S. Raby and S. Thomas, Phys.\ Rev.\
        Lett.\ 76, 3494 (1996); D. R. Stump, M. Wiest, C. P. Yuan, PHys.\
        Rev.\ D54, 1936 (1996); S. Ambrosanio, G. L. Kane, G. D. Kribs, S. P.
        Martin and S. Mrenna, Phys.\ Rev.\ Lett.\ 76, 3498 (1996); Phys.\
        Rev.\ D54, 5395 (1996); S. Dimopoulos, S. Thosmas and J. D. Wells, 
        Phys.\ Rev.\ D54, 3283; (1996);  
        K. S. Babu, C. Kolda and F. Wilczek, Phys.\ Rev.\ Lett.\ 77, 3070
        (1996);  K. Kiers, J. N. Ng and  G. Wu, Phys.\ Lett.\ B381, 177
        (1996); J. L. Lopez and D. V. Nanopoulos, Phys.\ Rev.\ D55, 4450
        (1997); H. Baer, M. Brhlik, C.-h. Chen and X. Tata, Phys.\ Rev.\
        D55, 4463 (1997); A. Ghosal, A. Kundu and B. Mukhopadhyaya, Phys.\
        Rev.\ D56, 504 (1997); D. A. Dicus, B. Dutta and S. Nandi, Phys.\
        Rev.\ Lett.\ 78, 3055 (1997); A. Ambrosanio, G. D. Kribs and 
        S. P. Martin, Phys.\ Rev.\ D56, 1761 (1997).
\bibitem{dubo} S. L. Dubovsky, D. S. Gorbunov and S. V. Troitsky, 
         hep-ph/9707357.
\bibitem{crs} C. Csaki, L. Randall and W. Skiba, Nucl.\ Phys.\ B479, 65
        (1996).
\bibitem{ellis} J. Ellis, G.B. Gelmini, J.L. Lopez, D.V. Nanopoulos and 
         S. Sarkar, Nucl.\ Phys.\ B373, 399 (1992).
\bibitem{lyth} D. H. Lyth and E. D. Stewart Phys.\ Rev.\ Lett.\ 75, 210
        (1995); Phys.\ Rev.\ D53, 1784 (1996).

\end{thebibliography}
\end{document}